\documentclass[11pt]{article}

\usepackage{cite}
\usepackage{epsf}

\def\bbox#1{\mbox{\boldmath$#1$}}

\def\corresponds{{\lower.2ex\hbox{=}}{\rm\kern-.75em^\triangle}}
\def\succsim{\succ\kern-.9em_\sim\kern.3em}
\def\precsim{\prec\kern-1em_\sim\kern.3em}
\def\slantfrac#1#2{\kern1em^{#1}\kern-.3em/\kern-.1em_{#2}}
\def\lfrac#1#2{{}^{#1\!}\kern-.0em/_{#2}}

\newcommand{\hf} {{1\over2}}
\newcommand{\nonu}{\nonumber\\}
\def\eq#1{(\ref{#1})}
\def\cd#1{{\cal D}[#1]}
\def\ord#1{{\cal O}(#1)}

\usepackage{amsfonts}

\begin{document}
\begin{center}
\begin{tabular}{c}
\hline
\rule[-5mm]{0mm}{15mm} {\huge \sf  
Renormalization of the Periodic Scalar Field Theory} \\ 
\rule[-5mm]{0mm}{15mm} {\huge \sf 
by Polchinski's Renormalization Group Method} \\
\hline
\end{tabular}
\end{center}
\vspace{0.2cm}
\begin{center}
I. N\'andori$^{1),2)}$, K. Sailer$^{2)}$,
U. D. Jentschura$^{1)}$,
and G. Soff$^{1)}$
\end{center}  
\vspace{0.2cm}
\begin{center}
$^{1)}$ Institut f\"ur Theoretische Physik,
Technische Universit\"at Dresden, \\
01062 Dresden, Germany \\[2ex]
$^{2)}$ Department of Theoretical Physics, 
University of Debrecen, \\
H-4032, Debrecen, Hungary 
\end{center}
\vspace{0.3cm}
\begin{center}
\begin{minipage}{11.8cm}
{\underline{Abstract} 
The renormalization group (RG) flow for the two-dimensional sine-Gordon model 
is determined by means of  Polchinski's RG equation at next-to-leading 
order in the derivative expansion. In this work we have two different goals, 
(i) to consider the renormalization scheme-dependence of Polchinski's method 
by matching  Polchinski's equation with the Wegner-Houghton equation and with 
the real space RG equations for the two-dimensional dilute Coulomb-gas,  
(ii) to go beyond the local potential approximation in the gradient expansion
in order to clarify the supposed role of the field-dependent wave-function
renormalization. The well-known Coleman fixed point of the sine-Gordon model 
is recovered after linearization, whereas the flow exhibits strong dependence 
on the choice of the renormalization scheme when non-linear terms are kept. 
The RG flow is compared to those obtained in the Wegner-Houghton approach and 
in the dilute gas approximation for the two-dimensional Coulomb-gas. }
\end{minipage}
\end{center}

\newpage

%\pacs{11.10.Hi, 11.10.Kk\\
%(Renormalization group evolution of parameters,\\
%Field theories in dimensions other than four)}

%
% Introduction
%
\section{Introduction} 

During the last three decades the renormalization of the two-dimensional 
Coulomb-gas or the equivalent sine-Gordon scalar model in dimension $d=2$ 
\cite{kerson} was investigated in a great detail either perturbatively 
\cite{col,mand,chle} or by means of the differential renormalization group 
(RG) method in momentum or in coordinate space, using a sharp 
\cite{per,coulomb} or a smooth cut-off \cite{kerson,kt,jkkn,wetter}. 
The well-known Kosterlitz-Thouless phase transition of the two-dimensional 
Coulomb-gas was obtained in the framework of the Kadanoff-Wilson 
renormalization scheme, using a dilute gas approximation 
\cite{kerson,kt,jkkn}. Several intuitive approaches exist by which one 
has tried to go beyond the dilute gas approximation 
\cite{coulomb,wetter,wieg,agg}. However, these attempts to  improve
the dilute gas approximation do not fit in a systematic scheme.

In this work we tried to go beyond the local potential approximation
in the gradient expansion. Our previous work \cite{coulomb} by means of
the Wegner-Houghton approach indicated that wave-function
renormalization may play an important role in the RG flow modifying it
for strong external fields. Furthermore, it showed that above a critical
value of the external field the path integral for the UV modes is
dominated by a non-trivial saddle-point and a spinodal instability occurs.
In order to clarify the supposed role of the wave-function renormalization 
we determine the RG flow for the periodic scalar field theory in dimension 
$d=2$ in this work. We discuss the field-dependent wave-function 
renormalization in the framework of Polchinski's RG method.

%
% The sine-Gordon, X-Y model and the Coulomb-gas
%           
\section{The sine-Gordon, X-Y model and the Coulomb-gas}

Several different models, like the sine-Gordon, Thirring, 
and the X-Y planar models belong to the same universality
class, namely to that  of  the two-dimensional Coulomb-gas. 
The X-Y model with an external magnetic field consists of 
classical two-component spins where the magnitude of the spin  is 
$\vert S_x\vert=1$ at each site. The model is given by the action:
\begin{eqnarray}
\label{XY}
S &=& {1\over T}\,\sum_{<x,x'>} \bbox{S}_x \cdot \bbox{S}_{x'} + 
{1\over T} \, \sum_x  \bbox{S}_x \cdot \bbox{h} \nonumber\\[2ex]
&=& {1\over T} \, \sum_{<x,x'>} \cos(\theta_{x}-\theta_{x'}) + 
{h\over T} \, \sum_{x} \cos(\theta_{x}) 
\end{eqnarray}
with the temperature $T$, the external field $h=|\bbox{h}|$, 
and the angle 
$\theta$ of each spin with an arbitrarily chosen direction. 
In the model there exist  topological excitations, called vortices,
which interact 
via Coulomb interaction. The X-Y model can be mapped by means of the 
Villain-transformation  to a Coulomb-gas \cite{kerson}. Such a mapping 
is, however, only valid up to irrelevant interaction terms. 
The other example, the two-dimensional sine-Gordon model is a 
one-component scalar field theory with periodic self-interaction, 
which is defined by the Euclidean action \cite{col}:
\begin{equation}\label{SG}
S = \int {\mathrm d}^2 x \left[{1\over2}\,(\partial \phi)^2 +
u\cos(\beta \phi)\right].
\end{equation}
with the Fourier-amplitude $u$ and the length of period $\beta$.
The equivalence between the X-Y model and the lattice regularized 
compact sine-Gordon model can be shown by expressing \eq{SG} 
in terms of the compact variable $z(x)=\exp[i\beta\phi(x)]$ \cite{kerson}.
This makes the kinetic energy periodic and introduces vortices in the 
dynamics.

%
% Polchinski's RG equation
%
\section{Polchinski's RG equation}

In  Polchinski's RG method \cite{polc} the realization of the differential 
RG transformations is based on a non-linear generalization of the blocking 
procedure using a smooth momentum cut-off. In the infinitesimal blocking
step
the field variable $\Phi(x)$ is separated into the slowly oscillating, IR 
($\phi$) and the fast oscillating, UV ($\tilde\phi$) parts, but both fields 
contain low- and high-frequency modes due to the smoothness of the cut-off. 
The propagator for the IR component is suppressed by a properly chosen 
regulator function $K(p^2/k^2)$ at high frequency above the moving
momentum 
scale $k$. The introduction of the regulator function generates infinitely 
many vertices with derivatives of the field. Most of these vertices are 
considered irrelevant and their flow is neglected. In order to determine 
 Polchinski's RG equation for the one-component scalar field theory,  
we follow here the method first explained in Ref. \cite{funcRG}. 

Let us start with the partition function for the scalar field $\Phi$ at 
scale $k$,
\begin{equation}
Z = \int \cd{\Phi} \hskip 0.1cm \exp\left(- S_k [\Phi]\right)
= \int \cd{\Phi} \hskip 0.1cm \exp\left(-{1\over2}\Phi G_k^{-1} \Phi 
- S_k^I [\Phi]\right),
\end{equation}
where the action is split into the sum of terms representing the
free propagation and the interactions,  
$\hf\,\Phi \, G_k^{-1}\, \Phi = (2\pi)^{-d} \int {\mathrm d}^d p 
\, \hf \, \Phi_{-p}\,G_k^{-1}(p^2)\,\Phi_p$
and $S_k^I[\phi]$, respectively, where 
$G_k^{-1}(p^2)=p^2 \, K^{-1}(p^2/k^2)$
stands for the regulated inverse propagator. The regulator function $K(z)$ 
suppresses the high-frequency modes ($|p| \gg k$) and keeps the 
low-frequency ones ($|p|\ll k$) unchanged due to the limiting behaviors 
$K(z) \to 0$ for $z \gg 1$ and $K(z) \to 1$ for $z\ll 1$, respectively.
Then the field variable and the propagator are split into the sum of IR
and UV parts,
\begin{equation}
\Phi= \phi + \tilde\phi,
\end{equation} 
and
\begin{equation}
G_k = G_{k -\Delta k} + {\tilde G_{k}},
\end{equation}  
where $\tilde G_k$ and $G_{k -\Delta k}$ correspond to $\tilde\phi$ 
and $\phi$, respectively. The UV propagator is written as 
$\tilde G_{k} = \Delta k \hskip 0.15cm \partial_k G_k$ for infinitesimal 
$\Delta k$. Since both $\phi$ and $\tilde\phi$ are non-vanishing for all 
momenta it seems as if the degrees of freedom were doubled. In order to
have 
the same number of degrees of freedom as  before the blocking one may
introduce 
a new dummy field $\tilde \Phi$ by inserting a trivial constant factor in
the partition function, written as a Gaussian path integral over the new field 
$\tilde \Phi$ with the arbitrarily chosen propagator $G_D (p^2)$,
\begin{eqnarray}
Z &=& \int \cd{\Phi} \hskip 0.1 cm \exp\left(-S_k [\Phi]\right)\nonumber\\[2ex]
&=& {\cal N} \int \cd{\tilde\Phi} \cd{\Phi}  
\hskip 0.1 cm \exp\left(-{1\over 2} \Phi \, G^{-1}_{k} \, \Phi 
-{1\over 2}{\tilde\Phi} \, {\tilde G^{-1}_{D}} \,{\tilde\Phi}- S_k^I
[\Phi]\right).
\end{eqnarray}
The fields $\phi$ and $\tilde\phi$ are defined by a 
Bogoliubov-Valatin transformation of the fields $\Phi$ and $\tilde\Phi$ 
\cite{funcRG} and the partition function becomes
\begin{equation}
Z=\int \cd{\tilde\phi} \cd{\phi} 
\hskip 0.1 cm \exp\left(-\hf\,\phi \, G^{-1}_{k -\delta k} \, \phi  
-\hf \, {\tilde\phi} \, {\tilde G^{-1}_{k}} \, {\tilde\phi}
- S_k^I [\phi+\tilde\phi]\right).
\end{equation}
The blocked action is defined by 
\begin{equation}\label{blocking}
\hskip 0.1cm \exp\left(-S^I_ {k-\delta k}[\phi]\right) = \int
D[\tilde\phi] 
\hskip 0.1cm \exp\left(-{1\over2} \tilde\phi \, {\tilde G_k^{-1}} \,
\tilde\phi 
- S_k^I [\phi+\tilde\phi]\right).
\end{equation}
We assume that the saddle point
\begin{equation}
  {\tilde \phi}_c = - {\tilde G}_k  {\delta S_k^I[\phi+{\tilde
      \phi_c}] \over \delta \phi}
\end{equation}
in this functional integral is 
$\ord{\Delta k}$ since $\tilde G_k=\ord{\Delta k}$. The expansion 
of the exponent around $\tilde\phi=0$ yields
\begin{eqnarray}
S_{k}^I [\phi]-S_{k-\delta k}^I [\phi]
& = & 
{1\over2}{\delta S_{k}^I [\phi]\over \delta\phi} \, {\tilde G_k} 
\, {\delta S_{k}^I [\phi]\over \delta\phi} \nonumber\\[2ex]
& & \quad - {1\over2} {\mathrm{Tr}} 
\ln\left[{\tilde G_k^{-1}} 
+ {\delta^2 S_{k}^I [\phi]\over \delta\phi\delta\phi}\right] 
+ \ord{\Delta^2k}.
\end{eqnarray}
Finally we perform the limit $\Delta k\to0$,
\begin{equation}\label{polchinski1}
\partial_k S_{k}^I [\phi]
={1\over2}{\delta S_{k}^I [\phi]\over \delta\phi} \, \partial_k G_k 
\, {\delta S_{k}^I [\phi]\over \delta\phi} -{1\over2} {\mathrm{Tr}}
\left[\partial_k G_k {\delta^2 S_{k}^I [\phi]\over 
\delta\phi\delta\phi}\right].
\end{equation}
One can rewrite \eq{polchinski1} for the complete action 
\cite{comellas,ball} as
\begin{eqnarray}\label{polchinski2}
\partial_k S_{k}[\phi]
&=& {1\over2} \, \int_p \partial_k G_k(p^2) \left[
{\delta S_{k}[\phi]\over\delta\phi_{-p}}{\delta
S_{k}[\phi]\over\delta\phi_p } 
- {\delta^2 S_{k}[\phi]\over \delta\phi_{-p}\delta\phi_p }
\right.
\nonumber\\[2ex]
& & \quad \left.
- 2\, \phi_p \, G_k^{-1}(p^2){\delta S_{k}[\phi]\over\delta\phi_p } \right].
\end{eqnarray}
The closed functional integro-differential
equation \eq{polchinski1} could only be obtained because the saddle point 
dominating the path integral of the blocking in \eq{blocking}
is $\ord{\Delta k}$. Such a suppression occurs because
the saddle point is driven by $S^I_k[\phi+\tilde\phi]$ and its 
amplitude is controlled by the kinetic energy
$\tilde\phi {\tilde G_k^{-1}} \tilde\phi/2$.
But it may happen that the kinetic energy is vanishing at a certain
momentum scale $k=k_c$ and a spinodal instability shows up.
One may think of models where the regulator function also evolves and
at given scale $k=k_c$, it starts to develop a singularity for some mode
$p_c$ for which $\left.{dK\over dp^2}\right|_{p_c}=\infty $. The saddle
point may become strong and an important tree-level renormalization
is observed in this case. This phenomenon is not covered by
Polchinski's equation and then a systematical search for the saddle 
point of \eq{blocking} has to be performed.

%
% Periodicity and Polchinski's equation
%
\section{Periodicity and Polchinski's equation}

In order to solve the functional integro-differential equations
\eq{polchinski1} and \eq{polchinski2}, one has to project them to a 
particular functional subspace.
It is generally assumed that the blocked action contains only local
interactions, and that
it can be expanded in the gradient of the field. We retain
the terms up to the next-to-leading order,
\begin{equation}
S_k = \hf \, \int {\mathrm d}^d x \, Z_k (\phi(x)) \hskip 0.1cm 
\partial_{\mu}\phi(x)\,\partial^{\mu}\phi(x)+
\int {\mathrm d}^d x V_k (\phi(x))
\end{equation}
with $Z_k (\phi)$ and $V_k (\phi)$ being the field-dependent wave-function 
renormalization and the potential. The interaction part of the action 
$S^{I}_k[\phi]$ is defined as
\begin{eqnarray}
\label{ansatz}
S^{I}_k[\phi] &=& S_k[\phi] - \hf \int\frac{{\mathrm d}^d p}{(2\pi)^d}
\, G^{-1}_k(p^2) \phi_p \phi_{-p}
\nonumber\\[2ex]
&=&  S_k[\phi] - \hf \int \frac{{\mathrm d}^d p}{(2\pi)^d} 
\, p^2 K^{-1} \left({p^2\over k^2}\right) 
\phi_p \phi_{-p} 
\nonumber\\[2ex]
&=& \hf \int\frac{{\mathrm d}^d p_1}{(2\pi)^d} \,
\frac{{\mathrm d}^d p_2}{(2\pi)^d} (-p_1 p_2) 
\left[Z_{-p_1 -p_2}(\phi) \right.
\nonumber\\[1ex]
& & \quad \left. - \hf \left(K^{-1}\left({p_1^2\over k^2}\right)
+K^{-1}\left({p_2^2\over k^2}\right) \right)\delta_{p_1+p_2} \right] 
\phi_{p_1} \phi_{p_2} + \int V_k(\phi)\,,
\end{eqnarray}
with $Z_{-p_1 -p_2}(\phi)= \int {\mathrm d}^d x \, Z_k(\phi_x)\, 
e^{-ix(p_1 + p_2)}$.
The Polchinski equation 
reduces to the RG equations for the dimensionless functions $Z_k(\phi_0)$
and 
$V_k(\phi_0)$ in dimension $d=2$,
\begin{eqnarray}
\label{polcNLO}
& & (2 + k\, \partial_k) V_k(\phi_0) = - [V^{(1)}_k(\phi_0)]^2 \, K'_0 +
(Z_k(\phi_0) - K^{-1}_0) \, I_1 + V^{(2)}_k(\phi_0) \, I_2\,, 
\nonumber\\[2ex]
& & k\,\partial_k Z_k(\phi_0) = -4 \,(Z_k(\phi_0)-K^{-1}_0) \,
V^{(2)}_k(\phi_0) K'_0  \nonumber\\[1ex]
& & \quad - 2 \, V^{(1)}_k(\phi_0) \, Z^{(1)}_k(\phi_0) \, K'_0  
- 2 \, [V^{(2)}_k(\phi_0)]^2 \, K''_0 + Z^{(2)}_k(\phi_0) \, I_2\,,
\end{eqnarray}
with $Z^{(n)}_k(\phi_0)=\partial^{n}_{\phi_0} Z_k(\phi_0)$,
$V^{(n)}_k(\phi_0)=\partial^{n}_{\phi_0} V_k(\phi_0)$, where 
$K'\equiv \partial_{{\tilde p}^2} K({\tilde p}^2)$,
$K'_0 = \partial_{{\tilde p}^2} K({\tilde p}^2)\vert_{{\tilde p}^2=0}$ ,
$I_1= (2\pi)^{-2}\int {\mathrm d}^2 {\tilde p} \, {\tilde p}^2 K'({\tilde p}^2)$,  
$I_2= (2\pi)^{-2}\int {\mathrm d}^2 {\tilde p} \, K'({\tilde p}^2)$ and 
${\tilde p}^2=p^2/k^2$. One can find the same RG equations for the
potential and for the wave-function renormalization in the literature by 
setting the anomalous dimension $\eta=0$ in Ref. \cite{comellas,ball}.

We consider the renormalization of the bare action  exhibiting the
symmetry
\begin{equation}
\label{pertr}
\phi(x) \to \phi(x)+\Delta,
\end{equation}
therefore the potential $V_k (\phi)$ and the wave-function renormalization
$Z_k (\phi)$ are expected to remain  periodic functions
 of the field with the length of period $\Delta$,
\begin{equation}
\label{perfc}
Z_k (\phi(x)) = Z_k (\phi(x) + \Delta), \hskip 1cm 
V_k (\phi(x)) = V_k (\phi(x) + \Delta).
\end{equation}
We shall furthermore assume that both the kinetic energy and the
interaction term of the action are periodic.
It can be seen that the Polchinski equation \eq{polchinski1} for the 
interaction part of the action keeps the periodicity of $S^I_k[\phi]$
with the unchanged period $\Delta$,
\begin{eqnarray}
S_{k-\Delta k}^I [\phi] & = &  S_k^I[\phi] \nonumber\\[1ex]
& & \quad + \Delta k\left(-{1\over2}{\delta
S_{k}^I[\phi]\over\delta\phi} \,\,
\partial_k G_k \,\, {\delta S_{k}^I [\phi]\over \delta\phi} +
{1\over2}\,{\mathrm{Tr}} \left[\partial_k G_k {\delta^2 S_{k}^I [\phi]\over
\delta\phi\delta\phi}\right]\right),
\end{eqnarray}
i.e. the energy $V_k(\phi)$ and the wave-function renormalization
$Z_k(\phi)$ satisfy \eq{perfc}.

Instead of  equation \eq{polchinski1}, which is valid for the
interaction 
part of the action, one can use the more usual form of the Polchinski
equation 
\eq{polchinski2} which is valid for the complete action. Although the
equation 
\eq{polchinski2} could appear to break the periodicity of the action  due to
the 
term $2\phi_p G_k^{-1}(p^2){\delta S_{k}[\phi]\over\delta\phi_p }$, 
this is not the case since
\begin{equation}
2\biggl( \phi_p +\Delta \delta_{p,0}\biggr) G_k^{-1}(p^2){\delta
S_{k}[\phi+\Delta]\over\delta\phi_p }
=2\phi_p G_k^{-1}(p^2){\delta S_{k}[\phi]\over\delta\phi_p }
\end{equation}
owing to the periodic nature of $S_k[\phi]$ and the property
$ G_k^{-1}(0)=0$.

%
% Linearized solution
%
\section{Linearized solution}

Here we consider the linearized Polchinski equations. First, we use
them to show that practically no constraints are laid upon the 
regulator function by requiring the same classification of the
scaling operators at the Gaussian fixed point as obtained in the
Wegner-Houghton method. Second, we solve the linearized equations
for the periodic blocked action and recover the Coleman fixed point.

One can linearize the equations \eq{polcNLO} around the UV Gaussian fixed 
point: $V_k(\phi_0)= V^* + \delta V_k(\phi_0)$ and  $Z_k(\phi_0)= Z^* + 
\delta Z_k(\phi_0)$  with $V^*=0$ and $Z^*=1$ (if the anomalous dimension 
$\eta$ is introduced, it is set to zero $\eta=0$). Then the linearized
equations are:
\begin{eqnarray}\label{polcNLOlin}
(2 + k\,\partial_k) \delta V_k(\phi_0) &=&  I_1 \, \delta Z_k(\phi_0) 
+  I_2 \, \delta V^{(2)}_k(\phi_0), 
\nonumber\\[2ex]
k\,\partial_k \delta Z_k(\phi_0) &=& 
-4(1-K_0^{-1}) \, \delta V_k^{(2)}(\phi_0) 
+ I_2 \, \delta Z^{(2)}_k(\phi_0).
\end{eqnarray}
%
%with $I_1= \int_{\tilde p} {\tilde p}^2 K'({\tilde p}^2)$ and 
%$I_2=\int_{\tilde p} K'({\tilde p}^2)$. 
In order to calculate the integrals $I_1$ and $I_2$, first one has 
to specify the regulator function $K({\tilde p}^2)$. One of the most 
important advantages of  Polchinski's RG method is the use of the 
smooth cut-off, which is compatible with the gradient expansion. 
Therefore it is possible to consider the renormalization of higher 
derivatives of the field, as well as the wave-function renormalization 
$Z_k(\phi)$. The price which has to paid, is 
that the RG equations depend on the particular choice of the regulator
function $K({\tilde p}^2)$. 
Unfortunately, this ambiguity cannot be easily removed. A rather
straightforward constraint on the regulator function $K({\tilde p}^2)$ is
that it should be defined in 
such a manner, that around the Gaussian fixed point the classification of 
the scaling operators into relevant and irrelevant ones be the same as 
that obtained with the Wegner-Houghton method. In order to formulate
this requirement, first we
rewrite
equation \eq{polcNLOlin} for the field-independent wave-function
renormalization
$\delta Z_k(\phi_0)= \delta z(k)$:
\begin{eqnarray}\label{polczlin}
(2+k\, \partial_k) \, \delta V_k(\phi_0) &=&  I_1 \, \delta z(k)+
I_2 \, \delta V^{(2)}_k(\phi_0), 
\nonumber\\[2ex]
k\,\partial_k \delta z(k) &=& -4\,(1-K_0^{-1})\,\delta V_k^{(2)}(\phi_0)  .
\end{eqnarray}
Then we can compare equation \eq{polczlin} to the linearized form of
the dimensionless 
Wegner-Houghton equation \cite{coulomb} around the Gaussian fixed point:
\begin{eqnarray}\label{whlin}
(2 + k\, \partial_k) \, \delta V_k(\phi_0) &=& - \alpha \,
\log\left[1+(\delta z(k)+ \delta V^{(2)}_k(\phi_0))\right]
\nonumber\\[1ex]
&=& - \alpha (\delta z(k) + \delta V^{(2)}_k(\phi_0))\,, 
\end{eqnarray}
and
\begin{equation}
k\,\partial_k \, \delta z(k) = 0\,,
\end{equation}
with 
\begin{equation}
\alpha=\frac{\Omega_2}{2(2\pi)^2}=\frac{1}{4\pi},
\end{equation}
where $\Omega_2=2\pi$ is the solid 
angle in dimension $d=2$. In order to get rid of the undesirable
tree-level
contribution in the linearized Polchinski equations, $K_0=1$ has to be
chosen. Then, both methods give no wave-function renormalization,
$z(k)=Z^*=1$. The comparison of the  field-dependent terms on the
right hand sides of   the first equations of \eq{polczlin} and
\eq{whlin}, we find the constraint $I_2 = - \alpha$, 
\begin{eqnarray}\label{int}
I_2 &=& \int {{\mathrm d}^2{\tilde p}\over (2\pi)^2} \, K'({\tilde p}^2) = 
\alpha \int_{0}^{\infty} {\mathrm d}x \, K'(x) = 
\alpha \, [K(\infty) - K(0)] =-\alpha, 
\end{eqnarray}
that is satisfied by any regulator function disappearing at infinity 
$K(\infty)=0$. Since the field independent terms on the right hands
side in the equations for the blocked potential depend on the
normalization, the value of the constant $I_1$ remains undefined.
There exist infinitely many regulator functions satisfying the above
mentioned rather weak constraints, an example is $K(x)= (1+a x^n)^{-1}$
or $K(x)= \exp(-a\,x^n)$.

The linearized equations \eq{polczlin} do not give an evolution for 
the field independent part of the wave-function renormalization 
$z(k)=1/\beta^2$, therefore we can rescale the field 
$\phi\rightarrow z^{-1/2}(k)\phi=\beta\phi$ 
and $\Delta={2\pi\over \beta}$ becomes the length of period
in the internal space, which remains unchanged during
the RG transformations. We can look for a solution of equation \eq{polcNLOlin} 
among periodic functions. The potential $\delta V_k(\phi_0)$ and the 
wave-function renormalization $\delta Z_k(\phi_0)$ are expanded in 
Fourier-series: 
\begin{eqnarray}
\delta V_k(\phi_0) =
\sum_{n=0}^{\infty} u_n(k) \cos\left(n\beta\phi_0\right),~~~~ 
\delta Z_k(\phi_0) = 
\sum_{n=1}^{\infty} z_n(k) \cos\left(n\beta\phi_0\right).
\end{eqnarray} 
For the sake of simplicity we consider the potential
 $\delta V_k(\phi_0)$ and the
wave-function renormalization $\delta Z_k(\phi_0)$ which possess the Z(2)
symmetry, 
$\phi_0 \leftrightarrow  -\phi_0$. The whole scale dependence occurs 
in the Fourier amplitudes $u_n(k)$ and $z_n(k)$, the `coupling constants' 
of the blocked action. The  linearized
 evolution equations for 
the Fourier-amplitudes read as follows,
\begin{eqnarray}\label{polclin}
(2 + k\,\partial_k) u_n(k) &=& 
I_1 \, z_n(k) -  I_2 \, u_n(k) \beta^2 n^2 ,\nonumber\\[2ex]
k\,\partial_k z_n(k) &=& - I_2 \, z_n(k) \beta^2 n^2 
\end{eqnarray}
for $n\ge 1$, where $I_2=-\alpha=-1/(4\pi)$ and the actual value of the 
integral $I_1$ is not fixed. 

For $Z_k(\phi)= z(k)$  independent of the 
field, the analytic solution
\begin{equation}
u_n(k)= u_{n0} \left({k\over \Lambda}\right)^{\alpha\beta^2 n^2-2},
{\mbox {  for  }} n\ge 0 
\end{equation}
exists in two dimensions with the initial values $u_{n0}$ at the UV
momentum 
cut-off $\Lambda$. This reproduces the well-known Coleman-fixed point and 
the  phases of the sine-Gordon model. Depending on the choice of $\beta^2$ 
the Fourier amplitude $u_n(k)$ is a relevant ($\beta^2 < 8\pi/n^2$), marginal 
($\beta^2 = 8\pi/n^2 $) or irrelevant ($\beta^2 > 8\pi/n^2$) coupling
constant. 

\begin{figure}[htb!]
\begin{center}
\begin{minipage}{10cm}
\centerline{\epsfysize=6.0cm\mbox{\epsffile{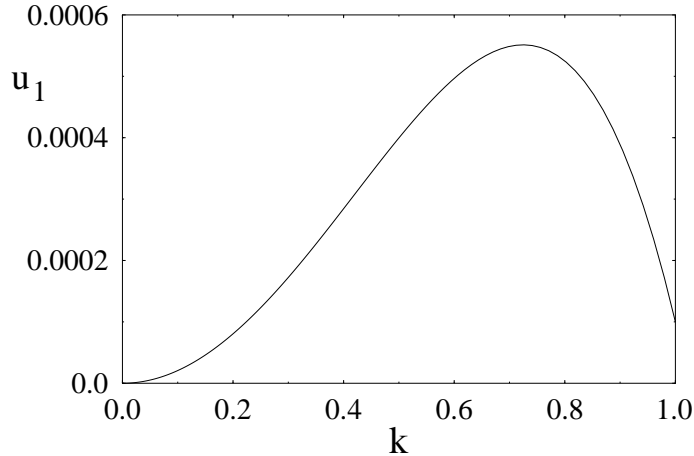}}}
\caption{\label{plot1}
The scaling of the Fourier-amplitude $u_1(k)$ is plotted
versus the running momentum cut-off $k$ from $k=1$ to $k=0$.
This result is obtained by solving the linearized form of Polchinski's RG
equation \eq{polclin} when $\beta^2=16\pi$ and the initial
value for the Fourier amplitudes are $u_1=0.0001$, $z_1=0.001 \beta^2$ and
$u_n=z_n=0$ if $n>1$ and the integrals $I_1=I_2=-\alpha$.}
\end{minipage}
\end{center}
\end{figure}

If the wave-function renormalization $Z_k(\phi)$ is field-dependent, 
then the second equation in \eq{polclin} can be solved analitically:
\begin{equation}
z_n(k)= z_{n0} \left({k\over \Lambda}\right)^{\alpha\beta^2 n^2},
{\mbox {  for  }} n\ge 1 . 
\end{equation}
Therefore, every $z_n(k)$ $(n\ge 1)$ is irrelevant whatever be the 
choice of $\beta^2$ and the actual value of the integral $I_1$. This
means that wave-function renormalization is irrelevant for both phases 
of the model in the UV regime, where the linearized equations hold. 
In this case, the solution for the Fourier amplitude $u_n(k)$ can be found 
only numerically, although in the IR domain, far from the UV cut-off 
$k \ll \Lambda$, the scaling of $u_n(k)$ only depends on the choice of
$\beta^2$, since all the coupling constants $z_n(k)$ are irrelevant. 
In Fig. \ref{plot1}, we plot the scaling of the Fourier amplitude $u_1(k)$
when $\beta^2=16\pi > \beta^2_c$ and the initial value 
for $z_1(k)$ is positive and the integral $I_1$ is set to be 
equal to: $I_1=I_2=-\alpha$. 
The position of the Coleman fixed point and the phase structure is
independent of the choice of the regulator fiction, that of the
renormalization scheme, while the actual flow depends on it.

%
% Fig. 1
%

%
% Comparison to the Coulomb-gas
%
\section{Comparison to the Coulomb-gas}

During the last two decades the real space RG approach for the Coulomb-gas 
in dimension $d=2$ has been investigated in great detail. The real space 
RG equations for the dilute vortex-gas are well-known and their derivation 
is given in the literature \cite{kerson,kt,jkkn}:
\begin{equation}\label{real}
a \, {dh(a)\over da}  = \left[ 2- \hbar \alpha T(a)\right]\, h(a), \hspace{1cm}
a \, {dT(a)\over da}  = - \hbar \, \pi \, T(a)^2 h(a)^2
\end{equation} 
with the lattice spacing $a$, and the dimensionless coupling constants $h$
and $T$. Due to the equivalence between the sine-Gordon model and the
two-dimensional vortex or Coulomb-gas, it is possible to compare Polchinski's 
RG equations for the sine-Gordon model with  equation \eq{real}. Since the 
real space RG equations for the Coulomb-gas are non-linear, one has to go 
beyond the linearized equations \eq{polcNLOlin} in order to compare the 
equations obtained by the two different methods.

In the real space RG equations \eq{real}  for the two-dimensional Coulomb-gas
only the field independent wave-function renormalization 
$Z_k(\phi_0)=z(k)=1/T=1/\beta^2$ is taken into account, therefore we
consider Polchinski's equations for the following two-dimensional model:
\begin{equation}
\label{sine-Gordon}
S_k[\phi] = \int {\mathrm d}^2 x \left[{1\over2}z(k)(\partial \phi)^2+u(k)
\cos(\phi)\right]
\end{equation}
where the two coupling constants are the Fourier amplitude
$u(k)=u_1(k)$ and the 
field independent wave-function renormalization $z(k)=z_0(k)$.
Inserting the ansatz \eq{sine-Gordon} in  Polchinski's equations
\eq{polcNLO}
and neglecting the terms on the right hand sides containing higher
harmonics,
we find
\begin{eqnarray}
\label{coulombpolc}
(2 + k\,\partial_k) u(k) = \hbar \, \alpha  u(k), \nonu
k \, \partial_k z(k) = - K''_0 \,  u^2(k).
\end{eqnarray}
The terms containing the derivatives of the wave-function 
renormalization with respect to the field $\phi_0$ do not appear in 
\eq{coulombpolc}, since $Z_k(\phi_0)=z(k)$ is independent of the field. In
the 
first equation of \eq{coulombpolc}, the field independent term $z(k) I_1$
and the term $[V^{(1)}_k(\phi_0)]^2 K'_0$  do not
give
contributions for the Fourier mode $\cos(\phi)$ since
$[V^{(1)}_k(\phi_0)]^2
= u^2(k) \sin^2(\phi_0)= u^2(k)(1/2 - 1/2 \cos(2\phi_0))$.
In the second equation of \eq{coulombpolc}, only $2 [V^{(2)}_k(\phi_0)]^2
K''_0$ 
gives field independent contribution, since
$[V^{(2)}_k(\phi_0)]^2= u^2(k) \cos^2(\phi_0)= u^2(k)(1/2 + 1/2
\cos(2\phi_0))$. 

These equations should be compared to the flow equations of the
sine-Gordon model obtained in the Wegner-Houghton approach
\cite{coulomb}, 
\begin{eqnarray}
\label{coulombwh}
(2 + k \, \partial_k ) u(k) = \hbar \,
\biggl[\alpha \frac{u(k)}{z(k)} + \ord{u^3}\biggr],
\nonumber\\[2ex]
k\,\partial_k z(k)
= -\hbar \, \biggl[{\alpha\over 2} \frac{u^2(k)}{z^2(k)} +\ord{u^4}\biggr].
\end{eqnarray}
These are Eqs. (7) of \cite{coulomb} rewritten for the dimensionless parameter
$u(k)$ and $z(k)$ when the higher order terms on
the right hand sides are neglected.
The significant difference is that the field independent wave-function
renormalization in \eq{coulombpolc} occurs due to
tree-level renormalization. Therefore,  the field-independent
piece of the wave-function renormalization depends on the `scheme',
which is equivalent to saying that it depends on the details of the 
blocking procedure (of Polchinski's type with various regulators, or of 
Wegner--Houghton's type). For the choice $K''_0=0$, and on the linear 
level, the scheme  gets closer to the WH approach as to the evolution of the
local potential, but the field-independent  wave-function renormalization 
does not alter during the blocking as opposed to the results obtained with 
the WH method, Eqs. \eq{coulombwh}.
Then the choice $z(k)=1/T(k) \equiv 1$ is consistent.

Using the equivalence between the lattice regularized compact 
sine-Gordon model and the X-Y model defined in \eq{XY},
\begin{equation}
z = 1/T = 1/\beta^2, \qquad  u = h/T\,,
\end{equation}
one finds that Polchinski's equations \eq{coulombpolc} can be
rewritten in the form of the two-dimensional Coulomb-gas as follows:
\begin{equation}
\label{coulombpolc2}
a \, {dh(a) \over da}  = \left[ 2 - \hbar \alpha T  \right] \, h,\qquad
a \, {dT(a) \over da}  = 0\, \rightarrow \,\,\, T(a)=T\equiv 1.
\end{equation}
These equations \eq{coulombpolc2} are rather different from those obtained
by the real space RG approach for the two-dimensional Coulomb-gas
\eq{real}. At the linear level the RG flow equations are identical
irrespectively of the blocking procedure by which they are obtained. The
differences of the various approaches occur when the non-linearities
 are kept that are responsible for the violation of the UV scaling laws.

%
% Summary
%
\section{Summary}

Our goal was to investigate how far the limitation of the
Wegner-Houghton approach due to the truncated gradient expansion 
can be overcome by using Polchinski's method combined with the 
gradient expansion. The price one has to pay for introducing the 
smooth cut-off and not clearly discriminating between UV and IR 
modes of the quantum fluctuations seems to be high.
It has been argued that the application of Polchinski's equations
to the renormalization of the scalar field theory with periodicity in
the internal space may be troublesome. First, the method mixes the
modes above and below the critical scale at which the spinodal
instability occurs. Therefore, if such an instability does exist as it is
expected from the investigations \cite{coulomb}
 by means of the Wegner-Houghton method,
 then the closer we come to this scale, the more
questionable 
the usage of the Polchinski's method is. Second, there is a regulator
dependent tree-level renormalization even if the path integral is
dominated by a trivial saddle point and no spinodal instability
occurs. The latter heavily depends on the choice of the regulator
function $K$, which is a priori not specified by any conditions.
Various forms of the cut-off function have been discussed in the 
literature; the choice was motivated by the need to reproduce the limiting 
behavior (critical exponents) \cite{comellas,ball}. Here, we take a 
different approach and we investigate if rigorous conditions can be 
established for $K$ based on the matching of Polchinski's equation 
with the Wegner-Houghton equation. The classification of the scaling 
operators at the Gaussian fixed point does not imply any substantial
contraints on the regulator function, but at least fixes the 
limiting behavior of $K$ at zero argument, and at infinity, 
in a unique way. Third, unfortunately, no 
further conditions can be obtained on the regulator function $K$ by 
comparing the dilute gas results with those obtained 
by Polchinski's method for the two-dimensional Coulomb-gas,
since the non-linear flow equations are rather different basically due
to the regulator-dependent tree-level renormalization in Polchinski's
approach.

Summarizing,  Polchinski's equation in its linearized form
 enables one to establish the Coleman fixed point and the phase
 structure of the two-dimensional sine-Gordon model. Furthermore,
 in the strong coupling phase $\beta>8\pi$ the linearization does not
 loose its validity in the limit $k\to 0$, and we have shown that 
 all couplings associated with the field 
dependent wave-function renormalization are irrelevant in this phase.
They are effectively `renormalized out' of the theory.
For the weak coupling phase the  flow equations in their non-linear
forms should be solved. Then a strong dependence of the flow on the
first and second derivatives of the regulator function at zero
momentum occurs, and the various renormalization schemes, 
Polchinski's,  Wegner-Houghton's, and the  real space ones
give rather different results, although all they were equivalent in
the linearized form. Polchinski's scheme gets closer to 
Wegner-Houghton's one if the derivatives of the regulator function at
zero momentum are vanishing. The dependence on the regulator
functions, and that on the renormalization scheme (the details of the
blocking procedure) modifies the
effective (blocked) couplings.

The scheme dependence of the blocked couplings is strongly related to
a more general question. When do we say that the more simple
theory reduced by the help of the RG is solved? What is the use of
knowing the evolution of the Wilson action? One may be inclined to say
that in the $k\to 0$ limit the blocked couplings become  physical
 observables, since only the single mode $p=0$ is kept. This is,
 however not true. First, the parameters of the bare (Wilson) action
are not observables, just quantities closely related to observables.
Second, the dynamics of the `last' mode is scheme dependent, similarly
to the dynamics obtained after eliminating any of the modes during the
subsequent blocking steps. This scheme-dependence can only be avoided
by using the effective action. The evolution equations for the
effective action describe real physics. Their solutions provide
 the values  of the one-particle irreducible (1PI) graphs and the
 observables are the graphs (transition amplitudes), not the couplings.    
According to this, the endpoint of these evolution equations
is the set of the exact 1PI Green functions. 

%
% Acknowledgments
%
\section*{Acknowledgments}

I.~N. and K.~S. thank J. Polonyi for useful discussions. 
This work has been supported by the projects 
OTKA T032501/00, NATO SA(PST.CLG 975722)5066, 
and M\"OB-DAAD 27 (323-PPP-Ungarn). I. N. is also 
grateful for support by DAAD. This work has also 
been supported by BMBF and GSI (Darmstadt).

\end{document}